\documentclass[preprint,showpacs,preprintnumbers,amsmath,amssymb,superscriptaddress]{revtex4}
\usepackage{graphicx}
\usepackage{dcolumn}
\usepackage{bm}
\usepackage{color}
\newcommand{\bvu}[1]{\ensuremath{\hat{\mathbf{#1}}}}
\newcommand{\bv}[1]{\mbox{\boldmath $\mathrm{{#1}}$}}

\begin{document}
\title{Monte-Carlo Simulations for model bent-core molecules
with fluctuating opening angle}
\author{W. J\'ozefowicz }
\email[e-mail address:]{wj503@york.ac.uk}
\affiliation{Department of Chemistry, University of York, Heslington, YO10 5DD York, United Kingdom}

\author{ L. Longa}
\email[e-mail address:]{lech.longa@uj.edu.pl}
\affiliation{Marian Smoluchowski Institute of Physics, Department of
Statistical Physics and  Mark Kac Center for Complex Systems
Research, Jagellonian University, Reymonta 4, 30-059 Krak\'ow, Poland}
\date{\today}
\begin{abstract}
We study the effect of fluctuations in opening angle of bent-core molecules on stability of the nematic phases.
The molecules are built out of two Gay-Berne centers of $\kappa=4$, $\kappa'=5$, $\mu=1$ and  $\nu=2$, each corresponding to one of the molecular arms.
Constant-pressure Monte Carlo (MC NPT) simulations are carried out for two versions of the model,  with fixed- and fluctuating  opening angle, where fluctuations are taken harmonic about the angle of $140^o$.
The systems studied are first cooled down from the isotropic liquid
to a highly ordered crystalline smectic phase and then heated up all the way back. The liquid crystalline phases found on cooling are
uniaxial nematic, biaxial nematic and crystalline, hexagonal biaxial smectic. On heating the biaxial nematic is superseded by the metastable crystalline phase.
Similar system with molecules of shorter arms ($\kappa=3$) was shown to exhibit a wide range of reduced density,
where the uniaxial nematic was absolutely stable [\emph{Liq. Cryst.} {\bf 29}, 483 (2002)]. For molecules with rigid arms of $\kappa=4$ and opening angles of  $125^o$, $130^o$,  $135^o$ and $140^o$
a  further broadening of the density range of the uniaxial nematic phase is observed. However, the
helical superstructures reported for $\kappa=3$ do not show up for $\kappa=4$.
\end{abstract}
\pacs{61.30.-v, 64.70.Md, 05.10.Ln}
\maketitle
\section{Introduction}

The most common phases found for bent-core liquid crystals are smectics \cite{Takezoe}. It has therefore been quite surprising to discover that these systems can also stabilize the thermotropic uniaxial ($N_U$)- and  biaxial nematic ($N_B$)  phases \cite{LAMadsen, BRAcharya, Prasad, Lehmann}, with quest for $N_B$ occupying  soft matter research for more than 30 years now ({\em see e.g.}\cite{Luckhurst-Nature}).
To date the factors crucial for absolute stability of the $N_B$ phase in bent-core systems are not fully understood.  In this work we carry out
MC NPT simulations for model bent-core (V-shaped) molecules to seek for the {\em minimal molecular
features} necessary to stabilize $N_B$.
An excellent recent review on theoretical- and computer simulation attempts to find
the $N_B$ phase in model systems is found in \cite{Berardi-review}. A brief summary of the results relevant for what follows is given below.

The lattice simulations with \emph{'bent-core'} dispersion centers occupying lattice sites of a cubic lattice  show that the $N_B$ phase can be  stabilized about the Landau point at, or close to, the 'tetrahedral' opening angle of  $cos^{-1}(-1/
3)=109.47^o$
\cite{Bates05, Bates06, Bates07}. Additional inclusion of the dipole-dipole interactions with dipole moments along the $C_2$ molecular axis can transform the Landau point into a Landau line \cite{Bates07}.

The presently existing simulations for model bent-core mesogens with centers of mass free to move
(off-lattice) are  less optimistic. The excluded volume effects (packing entropy) make the stable nematic phases, especially the biaxial one, rare to appear in these systems. For example, an ensemble of bend-core molecules formed by two rigid spherocylinders joined at one end stabilize only $N_U$ and isotropic fluid  \cite{Camp,Lansac}. Replacing spherocylinders with Gay-Berne (GB) centers of $\kappa=3$, $\kappa'=5$, $\mu=1$ and  $\nu=2$, also does not bring a stable biaxial nematic, as shown by  Memmer \cite{Memmer} and Johnston {\em et al} \cite{Johnston1,Johnston2}. Even the $N_U$ phase disappears for an intermediate opening angle of $170^o$.
Similar conclusions as concerning stability of $N_U$ can be drawn from NPT MC simulations of Dewar and Camp  \cite{Dewar} for the multi-site bent-core model of rigidly connected LJ spherical sites. Before proceeding further we would like to mention about an interesting observation in aforementioned simulations  of Memmer \cite{Memmer}  and Johnston {\em et al } \cite{Johnston1}. It concerns the appearance of a spontaneously chiral, helical state close to the uniaxial nematic-smectic phase transition.  Memmer,  however, has made it clear that the existence of the helical structure might be an artefact of the method he used for this structure was not reproduced in every run.

The Metropolis MC simulations for model bent-core molecules with dipole moments
have also been carried out with an attempt to bringing the molecular modeling close to the experimental situation. Unfortunately,  the presence of a single
dipole located on the $C_2$ molecular axis has led so far to neither $N_B$ nor $N_U$ \cite{Johnston2,Dewar1,SOrlandi}.
But two point dipoles placed at the centers of the lateral
sites of a three GB bent-core molecule were shown
to stabilize the $N_U$ phase in NPT MC simulations \cite{SOrlandi}.
Moving the dipoles towards the terminal positions of the
molecular arms broadened the uniaxial nematic
range. Finally, the molecules with terminal flexible chains also do not help to stabilize $N_U$ or $N_B$ \cite{Dewar1}, but the simultaneous
presence of flexible tails and molecular dipoles along the $C_2$ axis can lead to  the $N_U$ phase \cite{Dewar1}. However, in none of the off-lattice simulations mentioned so far the biaxial nematic phase was detected.

To date, the  only off-lattice simulation for bent-core systems that reveals the biaxial nematic phase has been that of Pel\'{a}ez and Wilson \cite{Pelaez06}.
They performed the Molecular Dynamics simulation for real molecules with experimentally detected $N_B$ phase  \cite{LAMadsen,BRAcharya} using a full-atomistic potential.
The biaxial nematic phase with local ferroelectric domains was observed on cooling down an isotropic phase. However, the same atomistic model
with electrostatic charges switched off gave a stable smectic phase instead of $N_B$. This seems to indicate  that a subtle balance between an average  bent
molecular shape, molecular flexibility and transverse dipoles is essential for bent-core mesogenic molecules to stabilize the $N_B$ phase, and that the question
of the minimal molecular features necessary to obtain $N_B$ is still open.

In this work we present results of Metropolis
MC NPT simulations aimed at seeking for such minimal molecular
features necessary to stabilize $N_B$ for V-shaped molecules.
We check as whether the  elongation of the molecular arms can play any role. We also study some consequences of fluctuations in opening angle. We choose the model similar to that  of Memmer \cite{Memmer}, where bent-core molecules were built out of  two Gay-Berne (GB) centers representing molecule's arms. An attractive feature of the model \cite{Memmer} is that it shows the stable $N_U$ phase in a  relatively wide temperature range. Our model differs from the original one  by longer molecular arms. In addition we allow for harmonic fluctuations in the opening angle. Both changes are shown to have a significant effect on the nematic polymorphism, indicating that the lengthening of the molecular arms can help stabilizing the $N_B$ phase.

The paper is organized as follows. In Sec. II we define model banana-shaped molecules and give some computational details.
In Sec. III we introduce relevant order parameters and distribution functions
used to identify structures.
Results are presented in Sec. IV and a short summary is given in Sec. V.

\section{Model and computational details}

\begin{figure}[htbp]
\centering
\resizebox{!}{0.18\textheight}{\includegraphics{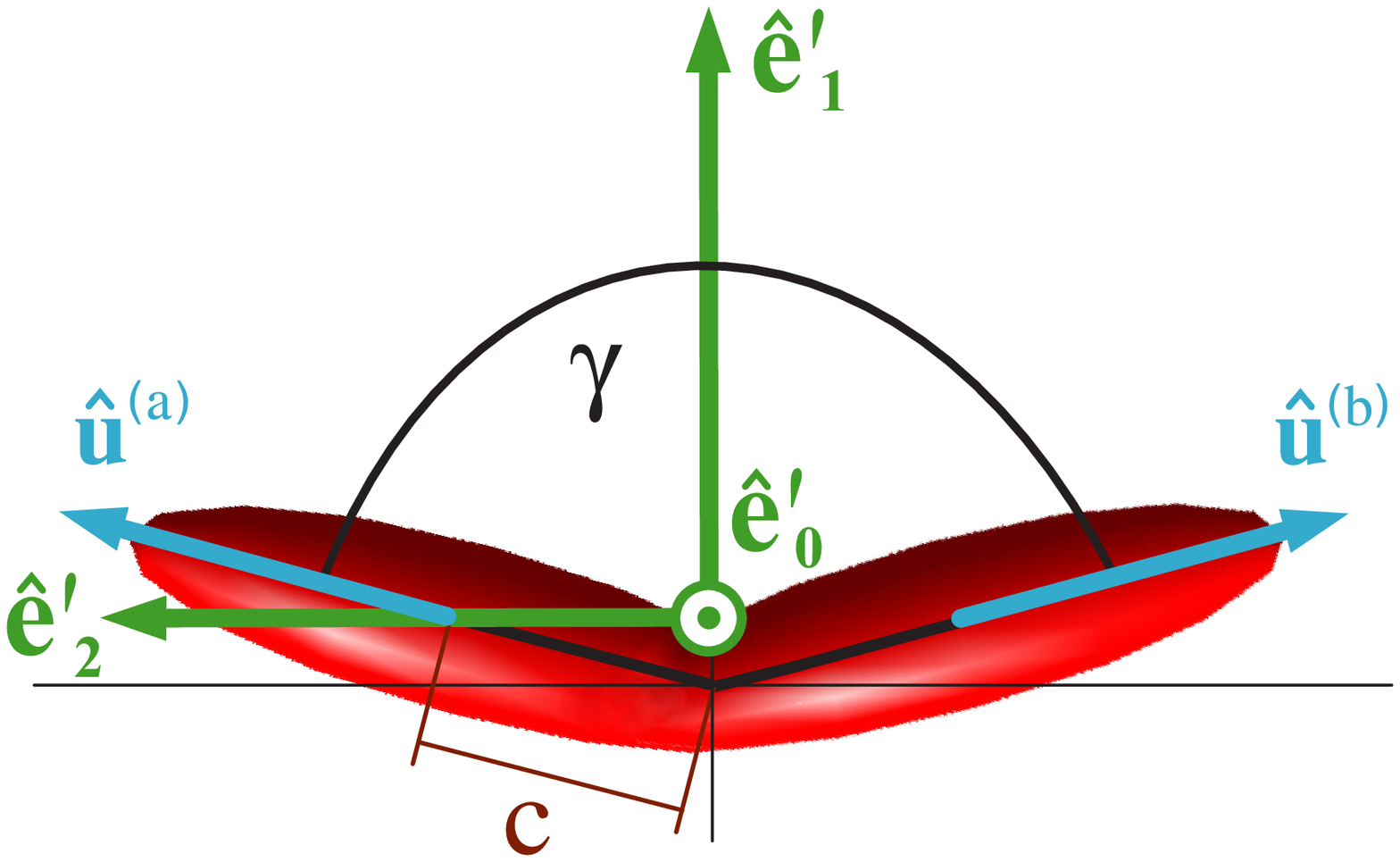}}
\footnotesize{\caption{\label{banan} Parametrization of model {bent-core} molecules used in simulations.}}
\end{figure}

We define a banana-shaped Gay-Berne molecule by linking two prolate
GB molecules, $a$ and $b$, as shown in Fig.~\ref{banan}. The positions of the GB centers  are
given by vectors $\bv{r}_a$ and $\bv{r}_b$.
In addition, each molecule is
parameterized with two internal coordinates: \emph{(i)} the opening angle
$\gamma$ ($0^0 < \gamma < 180^0$) between the  unit vectors $\bvu{u}_a$ and $\bvu{u}_b$ entering definition of the GB potentials
and (ii) the shift $c$ of the  GB centers  along the molecule's arms.
We also introduce a molecule-fixed orthonormal coordinate system with basis vectors $\{\bvu{e}{'}_\mu ,\,\,\, \mu=0,1,2 \}$, where $\bvu{e}{'}_2$ is taken
to belong to the line through endpoints of $\bv{r}_a$ and $\bv{r}_b$,
$\bvu{e}{'}_1$ is placed along the moleular $C_2$-axis, parallel to
$\bvu{u}_a+\bvu{u}_b$, and  $\bvu{e}{'}_0= \bvu{e}{'}_1 \times \bvu{e}{'}_2$.
Details of the
parametrization are given in Fig.~\ref{banan}.

The intermolecular interaction between a pair $(i,j)$ of the molecules,
separated by the intermolecular vector $\bv{R}_{ij}=\frac{1}{2}
[(\bv{r}_{a}^{(j)}+\bv{r}_{b}^{(j)})
-(\bv{r}_{a}^{(i)}+\bv{r}_{b}^{(i)})]$, is given by:
\begin{equation}
U(\{ \bvu{e}{'}^{(i)}_\mu \},\{ \bvu{e}{'}^{(j)}_\mu \},\gamma_i,\gamma_j,\bv{R}_{ij})=\sum_{\alpha=a,b}\sum_{\beta=a,b}
U_{GB}(\bvu{u}_{\alpha}^{(i)},\bvu{u}_{\beta}^{(j)}, \bv{r}_{\alpha
\beta}^{(i,j)}),
\end{equation}
where $\bv{r}_{\alpha \beta}^{(i,j)}=\bv{r}_{\beta}^{(j)}-\bv{r}_{\alpha}^{(i)}$ is
the intermolecular vector connecting the GB centers $\alpha,\beta$ ($\alpha,\beta=a,b$) of the molecules $i$
and $j$, respectively, with the vectors $\bvu{u}_i^{(\alpha)}$, $\bvu{u}_j^{(\beta)}$,  $\bv{r}_{ij}^{(\alpha \beta)}$
expressed through $\{\bvu{e}{'}_\mu^{(i)}\},\{\bvu{e}{'}_\mu^{(j)}\},\gamma_i,\gamma_j,\bv{R}_{ij}$
and $c$. Finally, the  $U_{GB}$ term is the Gay-Berne interaction \cite{GB} between the individual arms $\alpha$,
$\beta$ of the molecules $i$ and $j$, respectively. It reads
\begin{eqnarray}
U_{GB}(\bvu{u}_{\alpha}^{(i)}, \bvu{u}_{\beta}^{(j)},
\bv{r}_{\alpha \beta}^{(i,j)}) &=& 4\,
\epsilon(\bvu{u}_{\alpha}^{(i)}, \bvu{u}_{\beta}^{(j)},
\bvu{r}_{\alpha \beta}^{(i,j)})\,
\Bigg[
\left( \frac{\sigma_0}{r_{\alpha \beta}^{(i,j)}
-\sigma(\bvu{u}_{\alpha}^{(i)}, \bvu{u}_{\beta}^{(j)},
\bvu{r}_{\alpha \beta}^{(i,j)})
+\sigma_0}\right)^{12} \nonumber\\
&-& \left( \frac{\sigma_0}{r_{\alpha \beta}^{(i,j)}
-\sigma(\bvu{u}_{\alpha}^{(i)}, \bvu{u}_{\beta}^{(j)},
\bvu{r}_{\alpha \beta}^{(i,j)}) +\sigma_0}\right)^{6} \Bigg],
\end{eqnarray}
where $\bvu{r}_{\alpha \beta}^{(i,j)}=\bv{r}_{\alpha \beta}^{(i,j)}/r_{\alpha
\beta}^{(i,j)}$ is the unit vector and $r_{\alpha
\beta}^{(i,j)}=|\bv{r}_{\alpha \beta}^{(i,j)}|$.
The functions $\epsilon(\bvu{u}_{\alpha}^{(i)}, \bvu{u}_{\beta}^{(j)},\bvu{r}_{\alpha \beta}^{(i,j)})$
and $\sigma(\bvu{u}_{\alpha}^{(i)}, \bvu{u}_{\beta}^{(j)}, \bvu{r}_{\alpha \beta}^{i,j})$
depend on four GB parameters $\kappa$, $\kappa'$, $\mu$ and $\nu$ and their explicit form is found  in \cite{GB}. In addition,
each molecule $i$ is allowed to vary
its opening angle $\gamma_i$ about $\gamma_0$ with a harmonic contribution
to the potential energy
\begin{equation}
U_{int}^{(i)} = \left\{ \begin{array}{ll}
\infty & \textrm{\emph{if} $|\gamma_i-\gamma_0| > \gamma_{max}$} \\
k (\gamma_i-\gamma_0)^2 & \hspace{1cm} \textrm{\emph{otherwise}},
\end{array}\right.
\end{equation}
where $k$ is the harmonic constant, $\gamma_0$ is the angle for which internal energy has a minimum
and $\gamma_{max}$ is the angle of maximal deflection. The case of $k=\infty$ corresponds to a rigid molecule.

In our simulations we study the model with  the following GB parameters:
$\kappa=4$, $\kappa '=5.0$, $\mu=1$ and $\nu=2$. In addition we assume   $\gamma_0=140^o$, $\gamma_{max}=38^o$, $k=25, \infty$ and $c=1.7$,
where $c$ is chosen such that a smooth molecular shape is obtained
for opening angles in the range studied ($102^o - 178^o$).
Finally, throughout the paper angles are expressed in degrees
and all other parameters are rendered dimensionless.
Their definitions are found \emph{e.g.} in \cite{Memmer}.

Memmer \cite{Memmer} carried out MC simulations for a similar model, but with shorter molecular arms. The parameters different from ours were
$\kappa=3$ and  $c=1.0$. Only the case with rigid molecules ($k=\infty$) was considered. Thus, a comparison of our results with those of Memmer \cite{Memmer} will allows us to look at the influence of the arms' elongation and of the molecular arms fluctuations on the nematic polymorphism.

All simulations are performed using Metropolis MC simulations in NPT  ensemble
 \cite{Allen}.
The system consists of $N=1088$ molecules placed in a rectangular box
with periodic boundary conditions and the nearest image convention taken regard of.
A sequence of configurational states
is generated according to the probability density
\begin{eqnarray}\label{prawd}
f(\bv{\Gamma},\bv{\Omega},\bv{\Phi}) &=&
\Pi_{i=1}^N
\sqrt{\left(1+3\cos^2\frac{\gamma_i}{2}\right)\left(1-\cos^2\frac{\gamma_i}{2}\right)
\cos^2\frac{\gamma_i}{2}\left(4-3\cos^2\frac{\gamma_i}{2}\right)} \nonumber\\
&\,&\exp(-\,( \, U_{total}(\bv{\Gamma},\bv{\Omega},\bv{\Phi})+PV-NT \ln{V}\,
)/T \,  ) \,\,\, / \,\,\, Z^{'}_{NPT},
\end{eqnarray}
where $\bv{\Gamma},\bv{\Omega},\bv{\Phi}$ denote respectively
positions, orientations and opening angles  of all molecules, $V$
is the volume of the system, $U_{total}$ represents the total potential energy
and $Z^{'}_{NPT}$ denotes the partition function.
Boltzmann constant, not present in Eq.(\ref{prawd}),
is included into definition of the reduced temperature $T$.
The formula (\ref{prawd}) for the probability density  contains
square root factors coming from the integration over momenta. For rigid molecules the square root part gives a constant factor and, hence, cancels out with 
similar terms in $Z^{'}_{NPT}$.

The MC cycle  consisted of probing a
new configuration for each molecule followed by volume rescaling along a
randomly chosen edge of the simulation box. A
trial molecular configuration was generated by a random move of
the molecule's center of mass, a random rotation about randomly chosen axis out
of the laboratory reference system, a
flip of the steric dipole, and a random change of the opening angle.
In simulations $10\%-50\%$ of new molecular configurations
and $10\%-60\%$ of system rescalings were accepted. The lower acceptance ratios
were used for dense, highly ordered states, allowing to partly overcome the metastability traps.

Simulations were carried out under the pressure of $P=3$
and with spherical cutoff for the GB potential at the distance
$\bv{r}_{\alpha \beta}^{(i,j)}=5.0$.
The pressure was taken the same as in \cite{Memmer}.
Simulations were initialized in the isotropic liquid and
the temperature was lowered in adjustable steps until transition into
a highly ordered crystalline smectic phase took place. In addition,
the crystalline phase found on cooling was
heated up all the way back to estimate the stability range of the phases found.
The equilibration appeared slow and took between $5 \times 10^5$ and   $10^6$ cycles after which
the data  were collected for structure analysis once per 10 cycles over
a production run of $10^4$ cycles.

Interestingly, the biaxial nematic phase was found on cooling down.
To check its stability we also melted ground state crystalline structures under higher pressures, from  $P=4$ to $P=10$ in step of 1.

\section{Order parameters and distribution functions}
In order to characterize the long-range orientational order of the
molecules the alignment tensors were utilized. They are given by
\begin{equation}
Q^{\eta}_{\mu \nu}=\frac{1}{2N}\Big< \sum_{i=1}^{N}(3\,l_{\eta \mu}^{(i)} \, l_{\eta \nu}^{(i)} - \delta_{\mu \nu}) \Big>,
\end{equation}
where $l_{\eta \mu}^{(i)}=\bvu{e}{'}_\eta^{(i)} \cdot \bvu{e}_\mu$, and $\{ \bvu{e}_\mu \}$ constitute a base of the laboratory reference
system ($\eta, \mu, \nu \in \{0,1,2\}$).
Eigenvectors of  $Q^{\eta}_{\mu \nu}$ correspond to the directors of the phase
and the associated eigenvalues are the orientational order parameters measured with respect to these directors.
For isotropic phase all eigenvalues of each tensor $Q^{\eta}_{\mu \nu}$ vanish.
For uniaxial phase each tensor $Q^{\eta}_{\mu \nu}$ has one eigenvalue different than the other two, which are equal to each other.
In case of the biaxial phase all eigenvalues of  $Q^{\eta}_{\mu \nu}$
are, in general, different. However, it might happen that one of the tensors be uniaxial \cite{Averyanov}.

The presence of smectic order was detected by means of  the one-particle distribution function $P^{(1)}$, calculated along the main director. More specifically,
let the director $\bvu{n}$, corresponding to the largest eigenvalue of $Q^{2}_{\mu \nu}$,
be parallel to  $\bvu{e}_2$.
Then the smectic ordering along $\bvu{n}$ can be examined with the distribution function
\begin{equation}\label{banan_distr}
P^{(1)}(z)=\frac{\overline{L_z}}{N}\,\left\langle\,\sum_{i=1}^N\,
\delta(z-Z_i)\, \right\rangle, \enspace
\end{equation}
where $\overline{L_z}$ is the average length of the simulation box along
$\bvu{e}_2$, $Z_i$ represents coordinate of the $i$-th molecule along $\bvu{e}_2$,
and $\delta$ is the Dirac delta.

The in-layer structure of the smectic layers was examined with the two-point transverse correlation function
\begin{eqnarray}
g^{(2)}_\perp(r_\perp) &=&
\frac{\overline V}{4\pi r_\perp D  N^2 }\,
\left\langle\,\sum_{i}\,\sum_{j\neq i}\,
\delta(r_\perp-|{ \mathbf{R}}_i-{\mathbf{R}}_j|_\perp)\,
\Theta(D-|Z_i-Z_j|)\right\rangle,\hspace{4mm}
\label{banan_kor_p}
\end{eqnarray}
where $D=0.5$ is the thickness of the slice within which correlations were calculated,
$\overline V$ is the average volume of the simulation box, and $|{ \mathbf{R}}_i-{\mathbf{R}}_j|_\perp$
is the projection of the vector linking the origins $i$, $j$ of the molecule-fixed reference frames on a surface perpendicular to $\bvu{n}$. Owing the small system size we were unable to distinguish between hexatic- and crystalline orders and, for this reason,
a phase with long-range hexagonal order within layers is referred to as a crystal.

For flexible molecules the one-particle distribution function of opening angles was defined as:
\begin{equation}\label{banan_katy}
P^{(1)}(\gamma)=\frac{\pi}{N}\,\left\langle\,\sum_{i=1}^N\,
\delta(\gamma-\gamma_i)\, \right\rangle \enspace.
\end{equation}
Finally, the relative stability range for a given phase $f$ was examined by calculating the coefficients
\begin{eqnarray}
\Delta T^f = (T^f_H - T^f_L)/T^f_L, \\
\Delta \rho^f = (\rho^f_H - \rho^f_L)/\rho^f_H,
\end{eqnarray}
where $T^f_H$ ($\rho^f_H$), $T^f_L$ ($\rho^f_L$) represent upper and lower limit, respectively, of temperature (density),
for which the phase $f$ is detected.
\section{Results and discussion}
\begin{figure}[h!]
  \centering
  \vspace{-0.3cm}
  \begin{minipage}[t]{7 cm}
    \centering
    \includegraphics[scale=0.27]{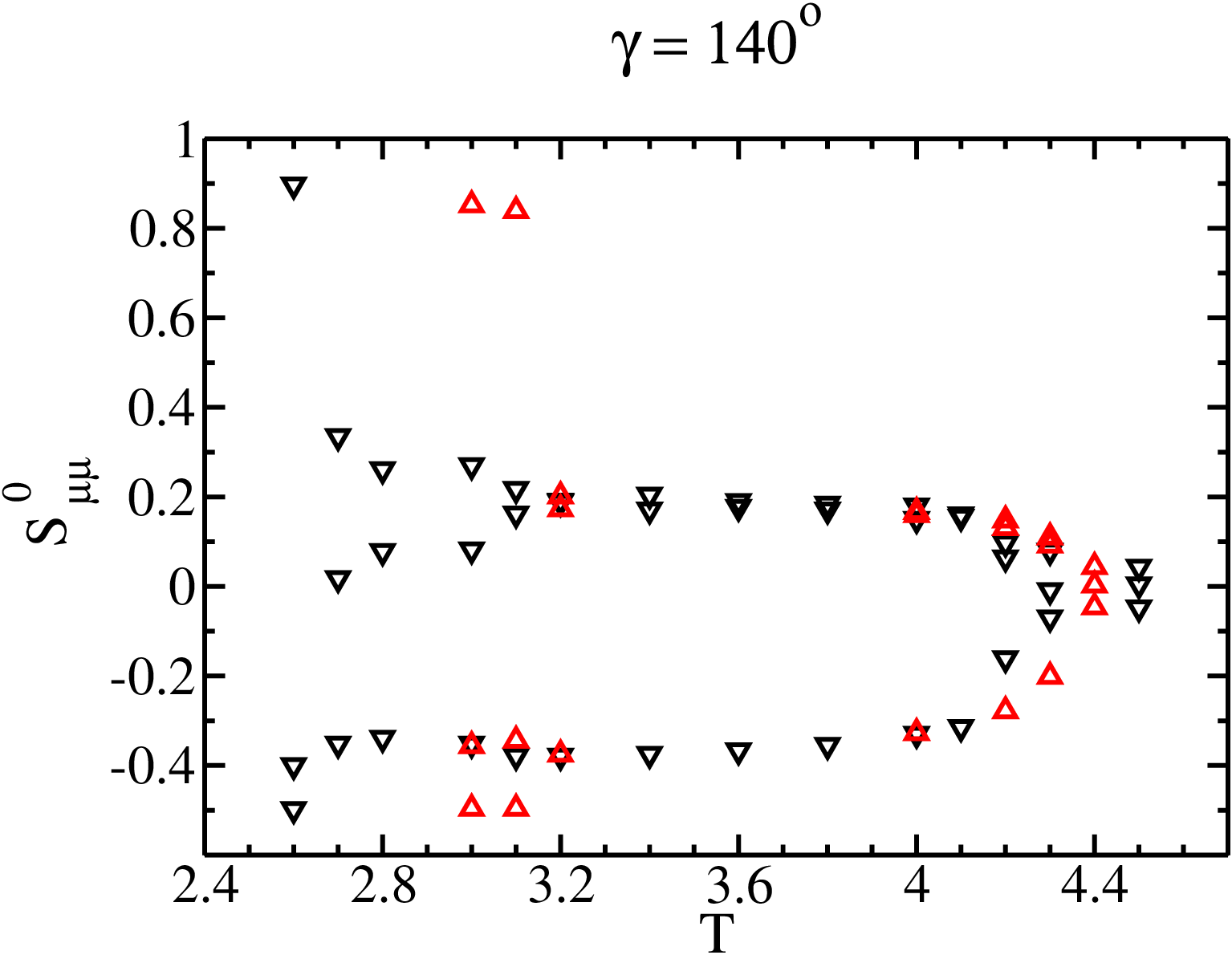}
  \end{minipage}
  \hspace{0.5cm}
  \begin{minipage}[t]{7 cm}
    \centering
    \includegraphics[scale=0.27]{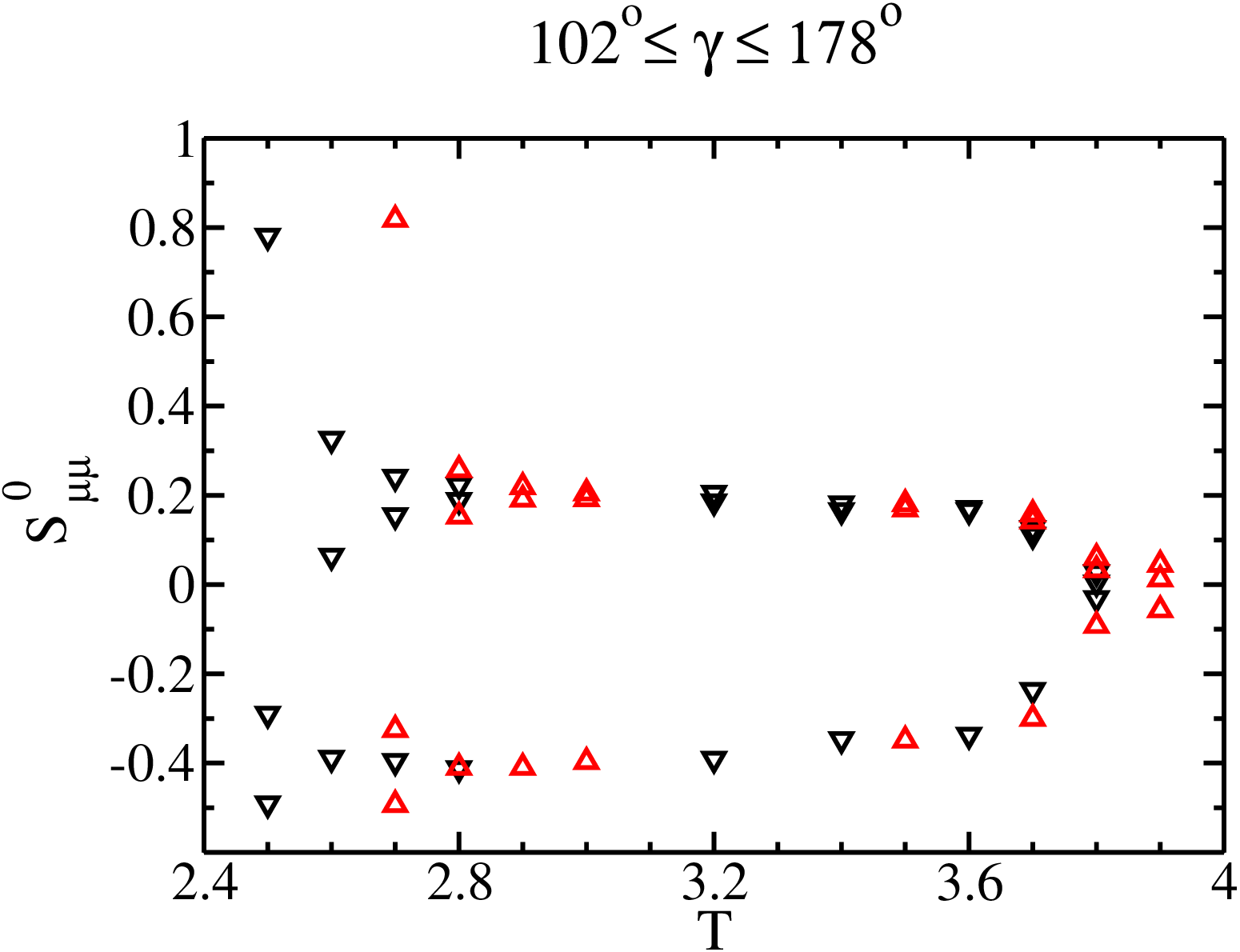}
   \end{minipage}
\end{figure}

\begin{figure}[h!]
  \centering
\vspace{-0.3cm}
  \begin{minipage}[t]{7 cm}
    \centering
    \includegraphics[scale=0.27]{140-const-Qy.eps}
  \end{minipage}
  \hspace{0.5cm}
  \begin{minipage}[t]{7 cm}
    \centering
    \includegraphics[scale=0.27]{140-var-Qy.eps}
  \end{minipage}
\end{figure}

\begin{figure}[h!]
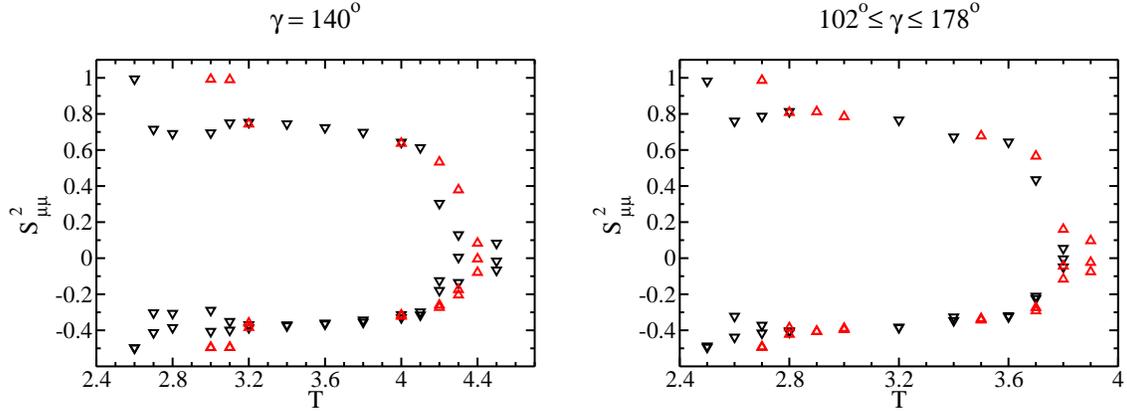

  \centering
\vspace{-0.3cm}
  \begin{minipage}[t]{7 cm}
    \centering
    \includegraphics[scale=0.27]{140-const-Qz.eps}
  \end{minipage}
  \hspace{0.5cm}
  \begin{minipage}[t]{7 cm}
    \centering
    \includegraphics[scale=0.27]{140-var-Qz.eps}
  \end{minipage}
\caption{\label{pic:Q-tens}
Eigenvalues $S^{\delta}_{\mu \mu}$ of tensors $Q^\delta_{\mu \nu}$
for system of rigid ($\gamma = 140^o$) and flexible ($102^o \leq \gamma \leq 178^o$) molecules obtained on cooling down
($\nabla$) and heating up ({$\triangle$}) the system.
The identified  phases are
a) orientationally isotropic, when eigenvalues of $Q^\delta_{\mu \nu}$ are close to zero;
b) uniaxial, when each tensor $Q^\delta_{\mu \nu}$ has only two different eigenvalues;
c) biaxial, when at least two tensors $Q^\delta_{\mu \nu}$ have all three eigenvalues different.
Note that the biaxial nematic phase appears when the system is cooled  down.
}
\end{figure}

\begin{figure}[h!]
\vspace{0.5cm}
  \centering
  \begin{minipage}[t]{7 cm}
    \centering
    \includegraphics[scale=0.27]{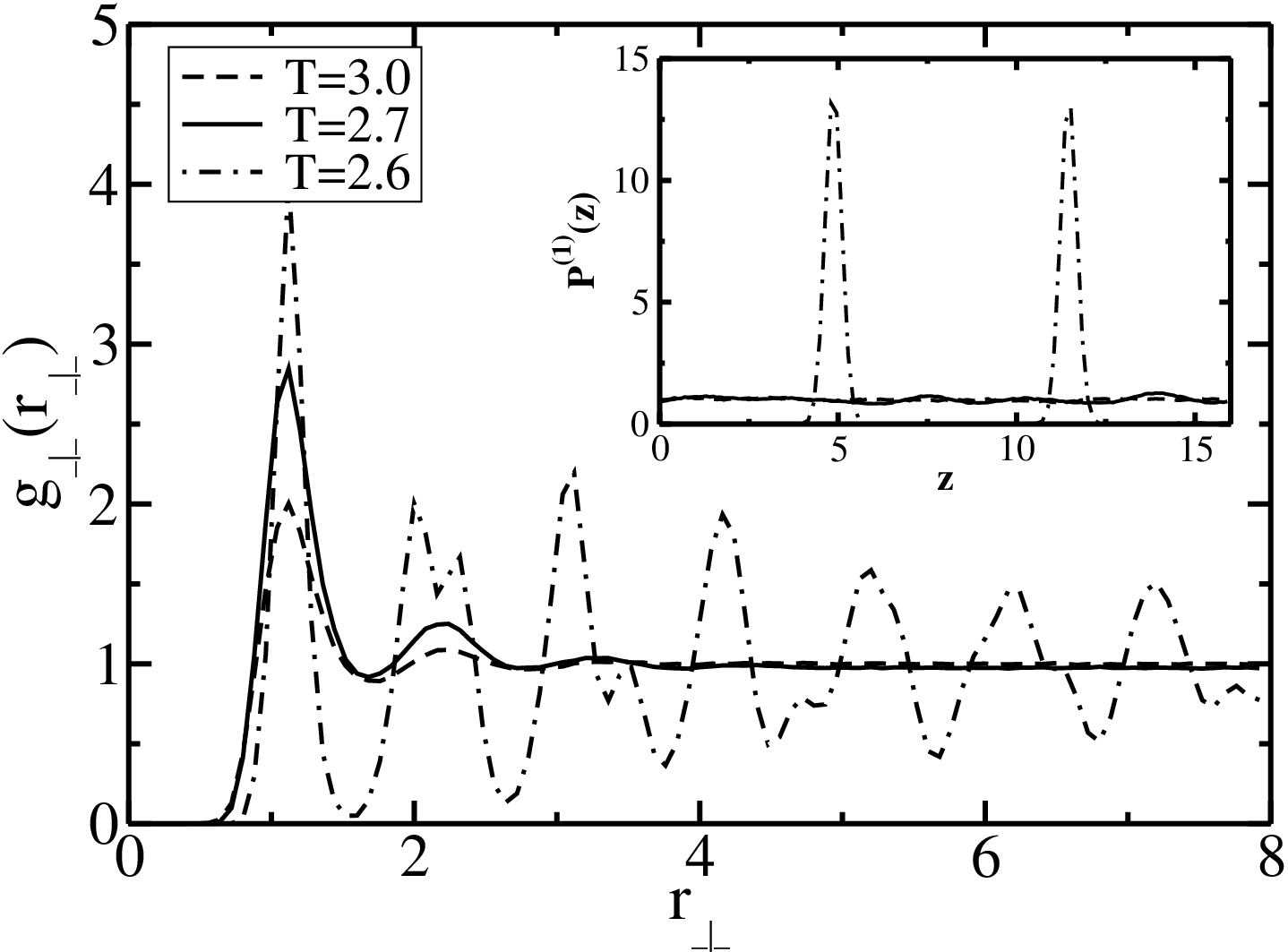}
   \caption{\label{pic:140-const-g}
Functions $g^{(2)}_\perp(r_\perp)$ and $P^{(1)}(z)$ obtained  for the system of rigid molecules (cooling down).
For $T \lesssim 2.7$ the system gains smectic order with hexagonal long-range order within layers.
}
  \end{minipage}
  \hspace{0.5cm}
  \begin{minipage}[t]{7 cm}
    \centering
    \includegraphics[scale=0.27]{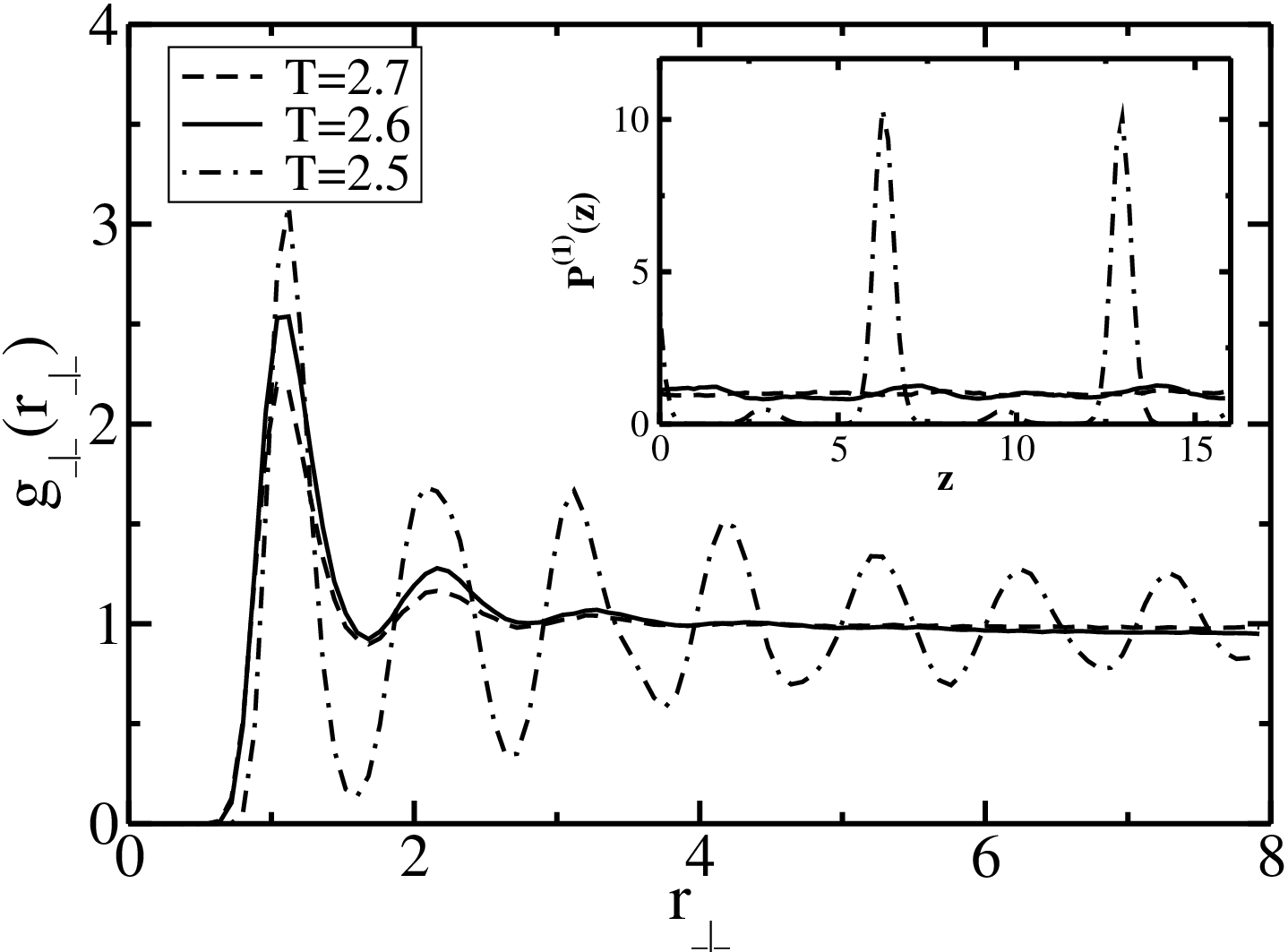}
    \caption{\label{pic:140-var-g}
Functions $g^{(2)}_\perp(r_\perp)$ and $P^{(1)}(z)$ obtained for the system of flexible molecules (on cooling).
For $T \lesssim 2.6$ the system gains smectic order with hexagonal long-range order within layers.
}
  \end{minipage}
\end{figure}

\begin{figure}[htbp]
\centering
\vspace{0.5cm}
\resizebox{!}{0.3\textheight}{\includegraphics{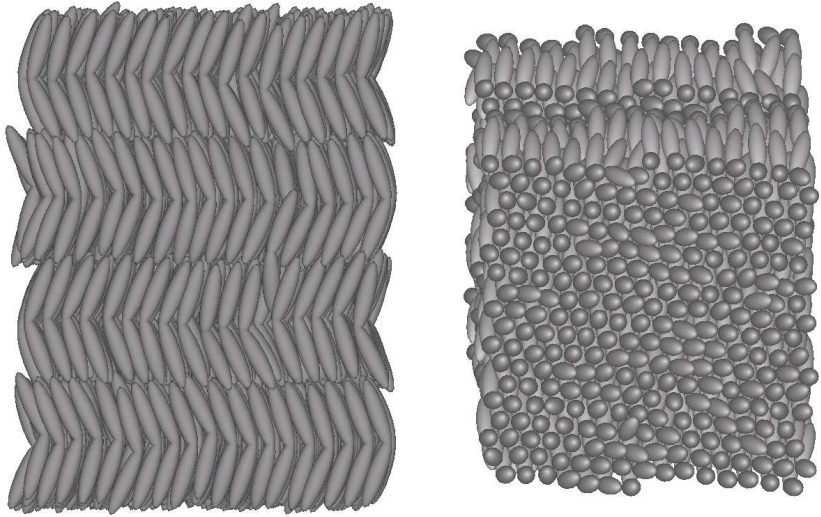}}
\footnotesize{\caption{\label{pic:banan-snap}
Left: side view of the crystalline phase obtained for the rigid model at   $T=2.6$.
Steric dipoles within neighboring layers have
opposite orientations making the crystalline structure ''antiferroelectric''.
Right: top view of hexagonal long-range order within smectic layers.
}}
\end{figure}

Cooling down the system of rigid molecules below $T\cong  4.1$ leads 
to a phase transition from isotropic liquid to uniaxial nematic (Figs.~\ref{pic:Q-tens},\ref{pic:140-const-g}).
Further lowering of the temperature causes stabilization of the biaxial nematic phase, which
is found for $2.7 \lesssim T \lesssim 3.0$ (Figs.~\ref{pic:Q-tens},\ref{pic:140-const-g}).
Periodic modulation of  $P^{(1)}(z)$ (Fig.~\ref{pic:140-const-g}) emerges only below $T\cong 2.8$,
but for $2.7 \lesssim T \lesssim 2.8$ this modulation is highly irregular, with amplitude not exceeding $0.2$.
Distinguishable smectic order is present only from below $T\cong 2.6$ and is accompanied by the hexagonal long-range order within
smectic layers (see $g^{(2)}_\perp(r_\perp)$ shown in Fig. \ref{pic:140-const-g}).
The phase is antiferroelectric with respect to the steric dipoles (Fig.~\ref{pic:banan-snap}). As we cannot distinguish between hexatic and crystalline order we classify this phase as a biaxial, hexagonal and antiferroelectric crystal.

The crystal melts into uniaxial nematic at  $T\cong 3.2$. On cooling down the biaxial nematic appeares
below  $T\cong 3.2$; hence without a credible calculation of the free energy it is not possible to settle if $N_B$ is really absolutely stable in our simulations (Figs.~\ref{pic:Q-tens},\ref{pic:140-const-rho}). Studies at higher pressures also do not resolve this issue unambiguously. However, owing that the hysteresis affects both measurements it is quite likely that the biaxial nematic might be absolutely stable in a small temperature interval.
Further increase of temperature transforms uniaxial nematic into isotropic liquid what takes place at $T \cong 4.4$.
The coexistence temperature of isotropic liquid and uniaxial nematic is pinpointed by localizing the discontinuity in the density
between these phases (Fig.~\ref{pic:140-const-rho}), which yields $T \cong 4.2$.

For the model bent-core molecules with shorter arms of $\kappa=3.0$ \cite{Memmer} no evidence of the biaxial nematic phase was reported.
Thus we conclude that molecules with more elongated arms help to stabilize $N_B$.
In addition, for Memmer's model
$\Delta T^N\cong0.5$ and $\Delta \rho^N\cong0.118$ while
 the same coefficients for our model  are
$\Delta T^N\cong0.313$ and  $\Delta \rho^N\cong0.127$. That is
the  elongation of the arms enhances the density range for which
nematic phases can be stable. Finally, in \cite{Memmer} a helical structure between uniaxial nematic and smectic has been reported.
Simulations we carried out do not show a formation of any helical order.

\begin{figure}[h!]
\vspace{0.5cm}
  \centering
  \begin{minipage}[t]{7 cm}
    \centering
    \includegraphics[scale=0.27]{140-const-rho.eps}
   \caption{\label{pic:140-const-rho}
Relation between density and temperature obtained for the system of rigid molecules
on cooling down ($\nabla$) and heating up ({$\triangle$}).
At $T\cong4.2$  isotropic liquid coexists with uniaxial nematic.
Hysteresis between liquid crystalline and crystalline phases
($T \leq 3.2$) is also shown.
}
  \end{minipage}
  \hspace{0.5cm}
  \begin{minipage}[t]{7 cm}
    \centering
    \includegraphics[scale=0.27]{140-var-rho.eps}
    \caption{\label{pic:140-var-rho}
Relation between density and temperature obtained for the system of flexible molecules
on cooling down ($\nabla$) and heating up ({}{$\triangle$}).
At $T\cong3.7$ isotropic liquid coexists with uniaxial nematic.
Shown is also hysteresis between liquid crystalline and crystalline phases
 ($T \leq 2.8$).
}

  \end{minipage}
\end{figure}

\begin{figure}[h!]
\vspace{0.5cm}
\centering
\resizebox{!}{0.24\textheight}{\includegraphics{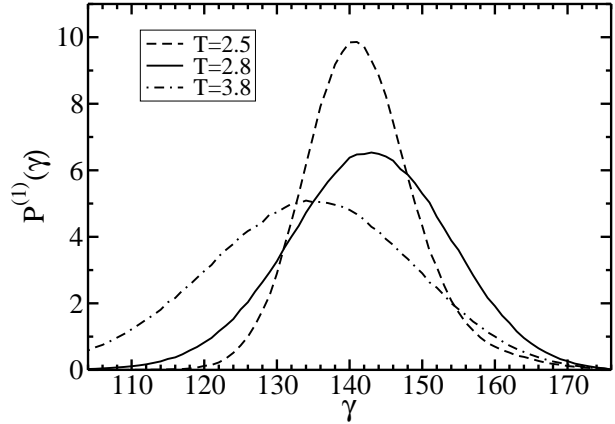}}
\footnotesize{\caption{\label{pic:140-var-katy}
Distribution of opening angles $P^{(1)}(\gamma)$ for system of flexible molecules.
}}
\end{figure}

Cooling down the system of the flexible bent-core molecules leads to the following phase transitions:
at $T\cong3.7$ the isotropic liquid transforms into the uniaxial nematic and at $T\cong 2.7$ the
nematic phase becomes biaxial (Figs.~\ref{pic:Q-tens},\ref{pic:140-var-g}).
For  $T\cong 2.6$ the one-particle distribution function, $P^{(1)}(z)$, develops an irregular modulation of magnitude smaller than 0.2. The ordering becomes periodic along the director for $T\lesssim 2.5$, which is attributed to the formation
of smectic layers.   Within the layers  the molecules form again a hexagonal lattice, Fig.~ \ref{pic:140-var-g},  with snapshots being similar
to the one shown in  Fig.~\ref{pic:banan-snap}. Hence,
the phase can again be classified as biaxial, hexagonal and antiferroelectric crystal.

The crystalline phase of the model melts directly into the uniaxial nematic phase at $T\cong2.8$  and, subsequently, the uniaxial nematic transforms  into
an isotropic liquid at $T\cong3.8$. From Fig.~\ref{pic:140-var-rho} we can estimate that the uniaxial nematic and the isotropic liquid coexist at $T \approx 3.7$. Like for the rigid model the presence of hysteresis (Figs.~\ref{pic:Q-tens},\ref{pic:140-var-rho}) does not allow to settle as whether the biaxial nematic is absolutely stable. Note, however, that fluctuations
in the opening angle seem to  have a
destabilizing effect on the biaxial nematic phase, as seen in Fig.~\ref{pic:Q-tens}.

For the model with flexible molecules the average opening angle, $\overline \gamma$,   changes with temperature (Fig.~\ref{pic:140-var-katy}).
At low temperatures,  in the crystalline phase, $\overline \gamma$   does not differ much from $\gamma_0$ (\emph{i.e.} $\overline \gamma\cong140.5^o$ for $T=2.5$). As temperature increases the average opening angle first increases to $\overline{\gamma} \cong 143^o$  in the uniaxial nematic phase ($T=2.8$) to next
decrease to $\overline{\gamma} \cong 135^o$  in the isotropic liquid phase ($T=3.8$).

The observation of $N_B$ on cooling encouraged us to perform  additional simulations at higher pressures. In these simulations we limited ourselves to the rigid model
of $\gamma_0=140^o$ and to integer values of pressure ranging from 4 to 14.
In addition, we carried out the simulations along the $P=10$ isobar for opening angles of $125^o, 130^o$ and $135^o$. In all cases we melted the ground state configurations and sought for an interval of stable biaxial nematic phase.
Apart from the model with the opening angle of $125^o$, the ground state crystalline structures melted directly to the uniaxial nematic phase. For the opening angle of $125^o$ we observed a direct melting to the isotropic liquid phase. As a result we cannot confirm that the biaxial nematic found on cooling is absolutely stable.

\section{Conclusion}

This paper contains results of MC NPT simulations performed for the system of model bent-core particles. The arms of V-shaped molecules were built out of two Gay-Berne sites.  Two variants of the model were studied, with rigid- and with fluctuating
angle between the arms. For both models decrease of temperature leads to the sequence of phases: isotropic liquid, uniaxial nematic, biaxial nematic
and biaxial antiferroelectric  hexagonal crystal. On heating the ground state configuration the biaxial nematic is superseded by a metastable crystalline phase.
However, in contrast to the model with molecules of shorter arms ($\kappa = 3.0$) \cite{Memmer}, the present systems, especially the rigid one, clearly show a tendency to stabilize $N_B$. In addition, the V-shaped molecules with longer arms extend the range of density in which the nematic polymorphism  can be present.
Finally, the helical structure found in \cite{Memmer}
was not detected in the present simulations.
\begin{acknowledgments}
This work was supported by Grant N202 169 31/3455 of the Polish Ministry of Science and Higher Education, and by the EC Marie Curie Actions 'Transfer of Knowledge', project COCOS (contract MTKD-CT-2004-517186).
\end{acknowledgments}

\end{document}